# Effects of excitation light polarization on fluorescence emission in two-photon light-sheet microscopy

GIUSEPPE DE VITO,[1,2] PIETRO RICCI,[2] LAPO TURRINI,[2,3] VLADISLAV GAVRYUSEV,[2,3] CAROLINE MÜLLENBROICH,[2,4,5] NATASCIA TISO,[6] FRANCESCO VANZI,[2,7] LUDOVICO SILVESTRI,[2,3,5] AND FRANCESCO SAVERIO PAVONE[2,3,5,*]

[1]University of Florence, Department of Neuroscience, Psychology, Drug Research and Child Health, Viale Pieraccini 6, Florence, FI 50139, Italy
[2]European Laboratory for Non-Linear Spectroscopy, Via Nello Carrara 1, Sesto Fiorentino, FI 50019, Italy
[3]University of Florence, Department of Physics and Astronomy, Via Sansone 1, Sesto Fiorentino, FI 50019, Italy
[4]School of Physics and Astronomy, Kelvin Building, University of Glasgow, Glasgow, G12 8QQ, UK
[5]National Institute of Optics, National Research Council, Via Nello Carrara 1, Sesto Fiorentino, FI 50019, Italy
[6]University of Padova, Department of Biology, Via Ugo Bassi 58/B, Padua, PD 35131, Italy
[7]University of Florence, Department of Biology, Via Madonna del Piano 6, Sesto Fiorentino, FI 50019, Italy

*francesco.pavone@unifi.it

**Abstract:** Light-sheet microscopy (LSM) is a powerful imaging technique that uses a planar illumination oriented orthogonally to the detection axis. Two-photon (2P) LSM is a variant of LSM that exploits the 2P absorption effect for sample excitation. The light polarization state plays a significant, and often overlooked, role in 2P absorption processes.

The scope of this work is to test whether using different polarization states for excitation light can affect the detected signal levels in 2P LSM imaging of typical biological samples with a spatially unordered dye population.

Supported by a theoretical model, we compared the fluorescence signals obtained using different polarization states with various fluorophores (fluorescein, EGFP and GCaMP6s) and different samples (liquid solution and fixed or living zebrafish larvae).

In all conditions, in agreement with our theoretical expectations, linear polarization oriented parallel to the detection plane provided the largest signal levels, while perpendicularly-oriented polarization gave low fluorescence signal with the biological samples, but a large signal for the fluorescein solution. Finally, circular polarization generally provided lower signal levels.

These results highlight the importance of controlling the light polarization state in 2P LSM of biological samples. Furthermore, this characterization represents a useful guide to choose the best light polarization state when maximization of signal levels is needed, e.g. in high-speed 2P LSM.

## 1. Introduction

Light-sheet (LS) fluorescence microscopy is a powerful optical imaging technique [1] based on the principle of a planar illumination oriented orthogonally with respect to the detection axis [2]. It employs wide-field detectors that allow to parallelize the photon collection, thus offering a large increment in the acquisition speed. Moreover, it offers also a good optical sectioning capability and reduced sample photodamage and photobleaching, compared to other optical imaging techniques [3].

Two-photon (2P) LS microscopy (LSM) [4–8] is a technique developed from traditional 1-photon (1P) LSM that exploits the 2P absorption effect for sample excitation [9]. The excitation wavelengths used in 2P absorption are usually in the infrared region: a frequency range characterized by reduced scattering inside biological tissues compared to visible light [10]. This effect, combined with the quadratic dependence of the absorption rate on the excitation light intensity, offers several additional advantages: a larger penetration depth in highly scattering samples [4], a reduction of the sample-induced aberrations, a better uniformity of the illumination distribution and an improved image contrast [11], without resorting to advanced illumination schemes [12–14]. Moreover, it allows the study of neuronal responses in conditions where they can otherwise be affected by the visible excitation light used in one-photon imaging [5,15,16].

The polarization state of the excitation light plays a significant, and often overlooked, role both in 1P and 2P absorption and emission processes exploited in microscopy, operating differently in the two cases [17], as we discuss in more detail in Section 2, "Theory". In particular, in the 2P absorption process, the sum of angular momenta of the absorbed photons is required to be zero, since the total angular momentum change related to the electronic state transition in most fluorophores is null [18,19]. Consequently, linearly-polarized light presents a higher 2P absorption than circularly-polarized light because it is favored with fulfilling this condition. On the other hand, circularly-polarized light will lead to a spatially more homogeneous fluorophore excitation, whereas in case of linearly-polarized light the excitation probability strongly depends on the relative orientation between fluorophores and beam polarization, leading to photoselection effects on the dyes [20].

It is well known that fluorescence emission from single dipoles is spatially anisotropic [21–23], because it happens preferentially on an axis perpendicular with respect to the emission transition moment [20], however this effect is usually averaged out when observing a large number of randomly-oriented rapidly rotating fluorophores. If the rotational diffusion rate is comparable to or slower than the fluorophore lifetime, then the photoselection effect induced by linearly-polarized excitation light will be preserved, at least partially, in the population of excited dyes, leading to anisotropy in the emitted fluorescence. The effects of this spatial anisotropy, as well as of the ellipticity of the excitation light, on biological imaging of randomly-oriented dyes were already experimentally characterized for 1P confocal microscopy and 2P microscopy [18] and they were studied also in other contexts such as fluorescence flow cytometry [24]. Nevertheless, their characterization is still lacking for fluorescent LSM, where the orthogonal geometry of the excitation and detection optical axes makes the presence of this anisotropy even more significant, as it was recently pointed out in the context of second harmonic generation LSM [25]. Since photoselection by polarized excitation can induce an anisotropically distributed emission pattern, it could be beneficial to experimentally orient the polarization orientation of excitation light to maximize the light emitted toward the direction of the detection objective in a LS microscope. In case there is an intrinsic spatial anisotropy in the fluorophore population orientation regardless of the excitation, i.e. when the dye population is spatially ordered due to the inherent biological properties of the sample, then polarization-resolved 1P or 2P fluorescence microscopy (as well as other microscopy techniques [26–30]) can in principle be used to extract information on the sample spatial-distribution or micro-architecture [31–35].

In this work we aim to phenomenologically test whether the use of different polarization states for the excitation light (circular polarization or linear polarization, with two orthogonal polarization orientations) can affect the detected signal levels when performing 2P LSM of biological samples in which the dyes are randomly oriented. The scope of the work thus is not to finely characterize the environmental factors responsible for this effect, but rather to highlight the importance of excitation polarization control in 2P LSM of biological samples and its use as a tool to increase the collected signal levels in this microscopy modality.

## 2. Theory

Fluorescence is a process resulting from two steps: the absorption of one (linear regime) or more (non-linear regime) incident photons followed by the incoherent emission of a single fluorescence photon, usually of lower energy. The emission incoherency entails that no phase correlation is preserved with the absorption step nor with the fluorescence radiated by other molecules. Consequently, the fluorescence intensity from a single molecule is proportional to the product between the n-photon absorption probability and the emission probability:

$$I^{n-ph} \propto P_{abs}^{n-ph} \cdot P_{em} \quad (1)$$

which greatly simplifies the process description, allowing to discuss separately the two contributions and the effects of polarization on them [20,27]. If the emission process maintains a phase relationship with absorption, as in second harmonic generation microscopy [25], some of the following considerations may need to be revised.

The n-photon absorption probability can be expressed as:

$$P_{abs}^{n-ph} \propto \left| \vec{\mu}_{ge} \cdot \vec{E}_{exc} \right|^{2n}, \quad (2)$$

where the dot denotes the scalar product between the polarized electric excitation field $\vec{E}_{exc}$ and the absorption transition dipole moment between the molecular ground state and the excited state $\vec{\mu}_{ge}$. The vectorial nature of the transition dipole moment can be intuitively understood as originating from the electron wave function distribution relative to the molecular structure and can be estimated using principles of quantum mechanics. Now, considering that the incident intensity is $I_{exc} \propto \left| \vec{E}_{exc} \right|^2$ and if we assume θ to be the angle between the polarization of the exciting electric field and the dipole moment of the molecule, it results that the 1P and 2P absorption probabilities are proportional to:

$$P_{abs}^{1-ph} \propto \left| \mu_{ge} \right|^2 I_{exc} \cos^2(\theta) \quad (3)$$

and

$$P_{abs}^{2-ph} \propto \left| \mu_{ge} \right|^4 I_{exc}^2 \cos^4(\theta), \quad (4)$$

respectively [20]. The dependency on the angle θ encodes the polarization effect, which is referred to as angular photoselection: absorption is most likely if the direction of the electric field of the photon(s) is aligned with the absorption transition dipole and no excitation will occur if the two are orthogonal, as depicted in Fig. 1. In the general case when these two vectors are not parallel, one can be decomposed into smaller parallel and orthogonal components in the coordinate system of the other, and only the first part will contribute to the absorption probability. Within a material (solid, liquid or gas) every molecule can be oriented differently, ranging from a co-aligned condition, as in a crystal, to a fully random angular distribution, as in a non-viscous liquid or ideal gas. The photoselection will entail that only the fraction of fluorophores with a non-zero projected component of their absorption transition dipole moment along the illuminating beam polarization direction will be excited, with a probability proportional to $\cos^{2n}(\theta)$ that steers towards narrower distributions for multi-photon processes, meaning that the photoselection effect is more pronounced in 2P microscopy with respect to 1P. The excitation beam can be either not polarized, entailing an isotropic excitation, or polarized linearly, leading to a cylindrically symmetric photoselection across the material, or polarized elliptically, in particular circularly. The latter two conditions can be described as a superposition of two orthogonal linear polarizations that varies spatially with period equal to the radiation wavelength, effectively inducing a more homogeneous excitation over the sample [17]. Another difference between these polarization types emerges in the 2P case where the sum of angular momenta of the absorbed photons is required to be zero, since the total angular momentum change related to the electronic state transition in most fluorophores is null [18,19]. It is more probable to fulfill this condition with linear rather that circular polarization, leading to a higher excitation rate, because in the former configuration there is a 50% probability for the fluorophore to interact with photons with opposite handedness that can reciprocally compensate their angular momenta.

The emission probability of a single photon along a certain observation axis $\hat{i}$ can be related to the radiated intensity $I_{em}(\hat{i})$, or equivalently to the field $\vec{E}_{em}(\hat{i})$ emanated by the molecular emission dipole moment $\vec{\mu}_{e'g}$ between an excited state (generally different and lower in energy than the one involved in absorption) and the ground state:

$$P^{em}(\hat{i}) \propto I^{em}(\hat{i}) \propto \left|\vec{E}^{em} \cdot \hat{i}\right|^2 \propto \left|\left(\hat{k} \times \left(\hat{k} \times \vec{\mu}_{e'g}\right)\right) \cdot \hat{i}\right|^2 = \left|\mu_{e'g}^{\perp} \cdot \hat{i}\right|^2 \quad (5)$$

where $\times$ denotes a vectorial product and $\hat{k} \times (\hat{k} \times \vec{\mu}_{e'g}) = \mu_{e'g}^{\perp}$ is the projection of the emission electric dipole moment on the direction orthogonal to the field propagation orientation $\hat{k}$. The latter relation represents the property that light waves (photons) travel in a direction perpendicular to their polarization sense. This means that the emitted field is partially polarized and is distributed anisotropically, presenting a cylindrical symmetry relative to the emission dipole moment orientation axis. Consequently, both the degree of polarization and radiation intensity will be highest in the plane orthogonal to this symmetry axis and zero in the parallel direction, as depicted in Fig. 1. The absorption and emission dipole moments are fixed relative to the molecular structure and they are parallel for many fluorophores, otherwise an additional angle between the two has to be considered.

Finally, combining the absorption and emission probabilities, we can conclude that the two-photon fluorescence intensity from a single molecule along the observation direction $\hat{i}$ is proportional to:

$$I^{2-ph}(\hat{i}) \propto \left|\mu_{e'g}^{\perp} \cdot \hat{i}\right|^2 \left|\vec{\mu}_{ge} \cdot \vec{E}_{exc}\right|^4 \propto \left|\mu_{e'g}^{\perp} \cdot \hat{i}\right|^2 \left|\mu_{ge}\right|^4 I_{exc}^2 \cos^4(\theta), \quad (6)$$

blending the excitation photoselection effects due to polarized illumination and the cylindrically symmetric output distribution related to the emission dipole moment. When many fluorescent molecules are present, their emission patterns will sum and the cumulative fluorescence intensity will depend on both their initial orientation during absorption and its eventually occurring evolution until emission (typical lifetimes $\tau$ are on the order of 1÷10 ns). Over this time interval, each molecule can undergo rotational diffusion, following a Brownian motion caused by thermal interaction with the environment [20]. This effect is inversely proportional to the anisotropy reduction factor:

$$\beta = \frac{1}{1 + \tau/\alpha}, \quad (7)$$

that is function of the ratio between the excited state lifetime $\tau$ and the rotational diffusion correlation time:

$$\alpha = \frac{\eta V}{K_b T} \quad (8)$$

that depends on the medium viscosity $\eta$, the temperature $T$ in Kelvin, the Boltzmann constant $K_b$ and the molecular volume $V$ of the dye or dye conjugate. Very small molecules in a dilute and low viscosity medium at room temperature can rotate very rapidly, with 10÷100 ps timescales, resulting in a completely randomized distribution that produces an isotropic emission. On the other hand, rotational diffusion is limited in viscous media, by large molecular size or by strong inter-molecular bounding to a much larger structure, such as in GFP, which damps their motional degrees of freedom, preserving partially or completely the initial anisotropy (either due to inherent anisotropic orientation or due to photoselection). Another process that may change and randomize the fluorophore arrangement is resonant energy transfer to other molecules [20]. In summary, the maximum possible anisotropy in emission will be observed when the distributions of absorbing and emitting dipoles are identical, requiring that their dipole moment orientations are parallel, do not rotate significantly during the excited state lifetime and no energy transfer between fluorophores occurs. Furthermore, the emission will be most intense when the excitation beam has matching linear polarization with the absorption dipole orientation. When one or more of these conditions are not respected, then the resulting emission pattern will tend towards becoming isotropic and less intense.

The impact of photoselection with Epi- or Trans-illumination and collinear detection has been widely studied [18,27,31–35], while fewer applications have been explored with orthogonal detection [24], as in the LSM geometry [25]. These detection configurations are not equivalent and the differences emerge when photoselection and polarized emission play an important role, while they become indistinguishable whenever the emission has an isotropic distribution, like when thermal diffusion dominates. Let us assume for the sake of clarity that all fluorophores have their dipole moment aligned along the camera detection axis, labeled as $\hat{z}$, while the laser beam propagation direction $\hat{x}$ and the orientation $\hat{y}$ orthogonal to both define the coordinate system, as illustrated in Fig. 1. The excitation light polarization will be defined as horizontal if parallel to $\hat{z}$ and vertical when along $\hat{y}$. In the first case, all fluorophores will be excited, but, due to the cylindrical symmetry of the dipole emission, they will emit mostly in the $xy$ plane and no light will reach the camera, as represented in Fig. 1(a). In the second case, no fluorophores will be excited due to the orthogonality between beam polarization and dipole moments, as shown in Fig. 1(b). If the dipole moments are instead oriented all along $\hat{y}$, as in Fig. 1(c), then a horizontal polarization will not excite the molecules, while a vertical polarization will produce the desired output prevalently in the $xz$ plane, reaching now the camera. It is interesting to observe that the camera will detect the same fluorescence distribution along all directions in the $xz$ plane due to the cylindrical symmetry of the emission pattern. Similarly, placing the camera in the $xy$ plane would lead to the same observations after swapping the two polarizations. On the other hand, if the camera is not in a generic position on these planes, but specifically on the $\hat{x}$ axis, as in Epi- or Trans-detection, then a signal will be observed for any polarization because only the beam propagation direction is shared by all intersecting planes. We would like to highlight that this is an important difference between LSM and Epi/Trans detection geometries. Finally, if the fluorophores are not aligned along an axis of the coordinate system, then the detected signal would depend on the observation direction. These considerations would apply also for an initially randomized fluorophore arrangement that emits partially polarized light due to photoselection, noting that the overall intensity will be generally lower than in the previous case, but more broadly distributed without reaching zero anywhere.

Summarizing, in the LSM geometry an excitation beam with polarization parallel to the imaging axis will lead to a generally lower measured fluorescence intensity than with orthogonal polarization and equality will be reached only for completely isotropic emission. This means that the polarization orientation of the excitation light could be experimentally oriented as to maximize the light emitted toward the direction of the detection objective in a LS microscope. We tested this prevision and the experimental results are reported in Section 4, "Results and discussion".

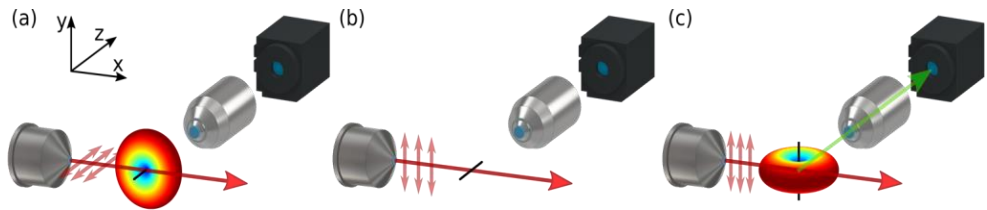

Fig. 1. Schematic of the polarization-dependent effects in 2P LSM, assuming a fixed orientation of the fluorophore dipole moment. (a) If both the polarization plane (indicated by dark-red arrows) of the linearly-polarized excitation light (red arrow) and the transition dipole of the fluorophore (black line) are aligned with the z-axis, the fluorophore is excited, but the fluorescence light (colored distribution) is emitted predominately on the xy-plane [24]. (b) If the polarization plane of the excitation light is parallel to the y-axis while the transition dipole is perpendicular to it, then no fluorescence light is generated. (c) If both the polarization plane of the excitation light and the transition dipole are aligned with the y-axis, then the fluorescence light is emitted predominately on the xz-plane and therefore part of it (green arrow) can be collected by the detection objective (along the z-axis).

## 3. Methods

We used two strains of transgenic zebrafish (*Danio rerio*) larvae: 3 Tg(elavl3:H2B-GCaMP6s) larvae [12,36] in homozygous *albino* background [37] and 6 Tg(actin:EGFP) larvae [38]. The former expresses, with nuclear localization, the fluorescent calcium sensor "GCaMP6s" under a pan-neuronal promoter, while the latter expresses enhanced GFP (EGFP) in all tissues owing to a ubiquitous promoter. Zebrafish strains were maintained according to standard procedures [39]. To avoid skin pigment formation, Tg(actin:EGFP) larvae were raised in 0.003% N-phenylthiourea (P7629, Sigma-Aldrich). All larvae were observed at 4 days post fertilization (dpf). Fish maintenance and handling were carried out in accordance with European and Italian law on animal experimentation (D.L. 4 March 2014, no. 26), under authorization no. 407/2015-PR from the Italian Ministry of Health.

Five of the larvae were subjected to live imaging. Immediately before the acquisition, each larva was anesthetized with a solution of tricaine (160 mg/L; A5040, Sigma-Aldrich), included in 1.5% (w/v) low gelling temperature agarose (A9414, Sigma-Aldrich) in fish water (150 mg/L Instant Ocean, 6.9 mg/L $NaH_2PO_4$, 12.5 mg/L $Na_2HPO_4$, pH 7.2) and mounted on a custom-made glass support immersed in fish water thermostated at 28.5 °C, as described in [40]. The other 4 larvae were fixed (2h in 4% paraformaldehyde in PBS at room temperature) before undergoing the same mounting procedure.

The imaging was performed with a custom-made 2P LS microscope. The setup scheme is shown in Fig. 2(a). Excitation light at 930 nm is generated by a pulsed Ti:Sa laser (Chameleon Ultra II, Coherent) and a pulse compressor is employed to pre-compensate for the group delay dispersion (PreComp, Coherent). The beam is attenuated using a half-wave plate and a Glan–Thompson polarizer and then it passes through an Electro-Optical Modulator used to rotate on command its linear polarization plane by 90°. Moreover, we use a combination of a half-wave plate and a quarter-wave plate to align the light polarization plane with the reference system of the microscope and to pre-compensate for the polarization distortions. The beam is then scanned by a fast resonant galvanometric mirror (CRS-8 kHz, Cambridge Technology), used to generate the digitally-scanned LS along larval rostro-caudal direction, while a closed-loop galvanometric mirror (6215H, Cambridge Technology) is used to scan the LS along larval dorso-ventral direction. The beam is finally relayed to an excitation dry objective (XLFLUOR4X/340/0,28, Olympus), placed at the lateral side of the larva, by a scan-lens (50 mm focal length), a tube-lens (75 mm focal length) and a pair of relay lenses (250 mm and 200 mm focal lengths) that underfill the objective pupil. When needed, we converted the light polarization state from linear to circular by placing a removable quarter-wave plate on the beam-path between the tube lens and the first relay lens.

The emitted green fluorescent light, coming either from GCaMP6s or EGFP, is collected by a water-immersion objective (XLUMPLFLN20XW, Olympus) placed dorsally above the larva. The objective is scanned along the axial dimension by an objective scanner (PIFOC P-725.4CD, Physik Instrumente) synchronously with the closed-loop galvanometric mirror movements. The optical image formed by the detection-objective tube lens (300 mm focal length) is then demagnified by exploiting a second pair of tube lens (200 mm focal length) and objective (UPLFLN10X2, Olympus), bringing the final magnification to 3×. Finally, the green fluorescence is spectrally filtered (FF01-510/84-25 nm BrightLine® single-band bandpass filter, Semrock) and relayed to a sCMOS camera (ORCA-Flash4.0 V3, Hamamatsu).

Imaging was performed with a pixel size of about $2\times2$ $\mu m^2$, and a field of view of about $1\times1$ $mm^2$. The acquisitions in fluorescein solution were performed on a single transversal plane with an exposure time of 100 ms. The larvae instead were imaged with volumetric acquisitions composed by 31 planes spaced by 5 μm and with an exposure time of 26 ms for each plane and a volumetric acquisition frequency of 1 Hz (~200 ms were reserved for objective flyback time). Each acquisition lasted 1 minute and then the 60 acquired volumetric stacks were averaged to obtain one final z-stack.

The laser power used for the acquisitions, measured at the excitation objective pupil, was 100 mW for the Tg(actin:EGFP) larvae both in living and fixed preparations, 200 mW for the live imaging of Tg(elavl3:H2B-GCaMP6s) larvae, 180 mW for the fixed Tg(elavl3:H2B-GCaMP6s) larvae and 162 mW for the fluorescein solution acquisition. Great care was taken to ensure that the excitation power remained constant when imaging with the three different polarizations. Moreover, we checked that this power range is far from the fluorescence saturation regime by measuring the average fluorescent signal generated by a fixed Tg(actin:EGFP) larva while varying the excitation power from 25 mW to 525 mW. The results, shown in Fig. 2(b), clearly depict a quadratic dependence of the signal from the excitation power (coefficient of determination: 0.999), as expected in 2P microscopy, and therefore we can exclude the presence of a saturation effect.

Before each acquisition session, we monitored the residual polarization distortions by temporarily inserting on the beam-path a half-wave plate followed by a Glan–Thompson polarizer before the excitation objective pupil. We then manually rotated the retarder while measuring the power variation after the polarizer. For circularly polarized light the amplitude of the observed oscillations was less than 4% of the signal.

General linear mixed models [41] were used to analyze the results for the Tg(actin:EGFP) larvae and the fixed Tg(elavl3:H2B-GCaMP6s) larvae. The models were implemented with the library "lmerTest" [42] for the R language for statistical computing. We used the fluorescent signal as dependent variable, the polarization state as fixed effect and the fish as random effect. A linear regression model implemented in R language was instead used to analyze the results for the living Tg(elavl3:H2B-GCaMP6s) larva. We used the fluorescent signal as dependent variable and the polarization state and the Region Of Interest (ROI) as independent variables. In both cases we used linear contrasts to compare the polarization groups and we used the Sidak method for the multiplicity correction. Fluorescein solution data were compared by computing 95% Confidence Intervals (C.I.) using the Student's t-distribution. In the following, all the fluorescence signal values are expressed in Arbitrary Units (A.U.).

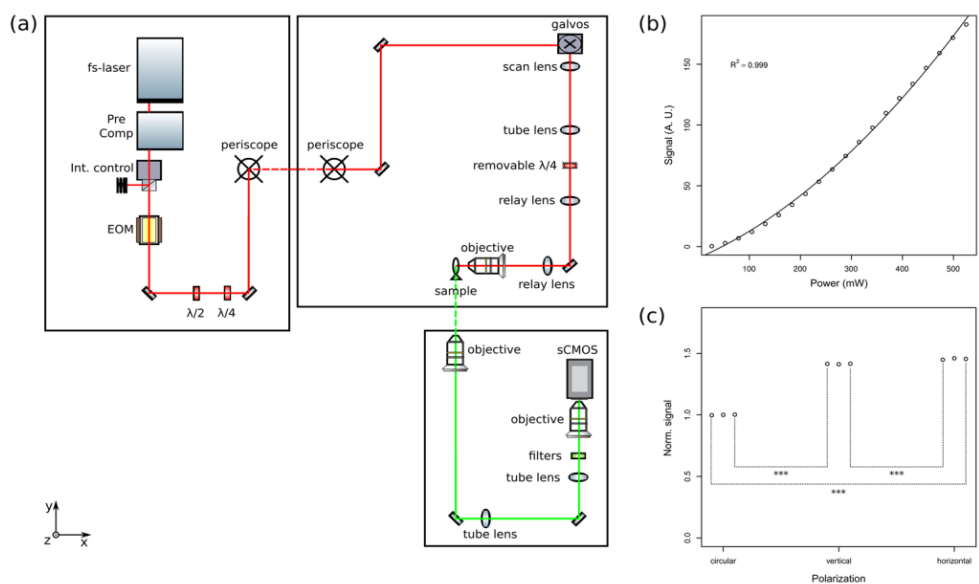

Fig. 2. (a) Schematic of the custom-made 2P LS microscope. Fs-laser: femtosecond laser. Pre Comp: pulse compressor. Int. control: intensity control assembly, composed by a half-wave plate and a Glan–Thompson prism. EOM: Electro-Optical Modulator. $\lambda/2$: half-wave plate. $\lambda/4$: quarter-wave plate. Galvos: galvanometric mirror assembly, composed by a resonant mirror and a closed-loop mirror. Red line: excitation light. Green line: fluorescence light. The dashed lines indicate vertical paths. Axis specification is consistent with Fig. 1, with the excitation and detection objectives oriented along the x-axis and the z-axis, respectively. (b) Scatter plot of the fluorescent signal generated by a fixed Tg(actin:EGFP) larva as a function of the excitation power. Parabolic fit of the data is indicated by the continuous line and its coefficient of determination is reported on the graph. (c) Scatter plot of the signal generated by a fluorescein solution excited with circularly-, vertically- or horizontally-polarized light. Each condition was tested in triplicate and each point represents a single measure. The signal value is normalized to the average of the circular polarization case. Statistically significant differences are indicated by three asterisks.

## 4. Results and discussion

By exciting small molecular-weight fluorophores in a medium where they are able to rotate completely unrestrained, we would expect the fluorescence emission to be isotropic, because in this condition the thermally induced rotation movements happen on time scales much shorter than fluorescence lifetime, washing out all anisotropy induced by photoselection, as discussed in Section 2.

To test this prediction, we excited fluorescence in a high-concentrated fluorescein solution employing linearly- and circularly-polarized light, and we show the results in Fig. 2(c). The polarization plane of the former was aligned either parallel or perpendicular to the plane where the optical axes of the detection and excitation objectives lay; in the following we adopt the convention introduced in Section 2 and refer to the parallel condition as "vertical polarization" and the perpendicular condition as "horizontal polarization" (corresponding to the z-axis and the y-axis, respectively, in Figs. 1 and 2).

We indeed observed similar fluorescence levels when employing vertically- or horizontally-polarized light, nevertheless we revealed a small, albeit statistically significant, difference between the two polarization states: the signal in horizontal-polarization condition (544.7 A.U.; 95% C.I.: [539.2, 550.2] A.U.) is ~3% larger with respect to the vertical-polarization condition (529.5 A.U.; 95% C.I.: [527.6, 531.5] A.U.). This observation indicates that even in a situation that favors high-level of molecular mobility (i.e. a small fluorophore in solution) a residual degree of spatial anisotropy induced by photoselection can be detected in fluorescence emission, with increased signal for horizontal polarization since it fulfills the orthogonal alignment condition between the excitation polarization direction and the imaging axis, as expected for the LSM detection geometry. A much larger difference was instead observed for circularly-polarized light: for this condition, we observed a ~30% reduction in the fluorescence signal level (374.4 A.U.; 95% C.I.: [372.0, 376.8] A.U.) with respect to the two linear polarization conditions. This is consistent with the expected lower 2P excitation efficiency that characterizes circular polarization. Nevertheless, it is not a trivial result, since circular polarization can excite the dyes in a spatially-homogeneous fashion, while the linearly-polarized light can excite only the subset of dyes that have a component parallel to its polarization-plane, as discussed in Section 2. This result therefore indicates that in the case of isotropic emission the widening of the group of possible target dyes is not sufficient to compensate the decrease in excitation efficiency for the circular polarization with respect to the linear polarization.

We then tested if this polarization-dependent effect is present also in tissue imaging and if the constraints introduced on the fluorophores by being embedded in the tissue structure would lead to a higher degree of photoselection-induced anisotropy in emission. To do so, we observed zebrafish larvae expressing EGFP, both in fixed and in living conditions, and we show the results in Fig. 3. In this case, we selected an arbitrary ROI for each larva (as depicted in Figs. 3(a) and 3(b)) and we measured its mean fluorescence signal. We did not observe significant differences between the circular-polarization condition and the vertical-polarization condition in both the fixed (41.6 A.U., standard deviation: 15.3 A.U. and 45.0 A.U., standard deviation: 14.2 A.U., respectively) and the living (53.4 A.U., standard deviation: 18.8 A.U. and 48.9 A.U., standard deviation: 17.5 A.U., respectively) conditions. We observed instead a large and significant ($p$-value $<$ 0.0001) signal increase in the horizontal-polarization condition with respect to the circular- and the vertical-polarization conditions, both in the fixed-condition (~67% and ~54%, respectively; horizontal-polarization value: 69.3 A.U., standard deviation: 20.4 A.U.) and in the living condition (~41% and ~54%, respectively; horizontal-polarization value: 75.2 A.U., standard deviation: 29.2 A.U.).

The presence of signal for all tested polarizations indicates that emission is not fully polarized and that there is a significant degree of randomness in the fluorophore orientation in the animal tissue, otherwise no fluorescence should have been observed for vertically polarized excitation, as noted in Section 2 for the LSM detection configuration. On the other hand, the difference in the signal levels between the horizontal- and the vertical-polarization conditions is much more pronounced than in the fluorescein solution. This distinction could be ascribed primarily to the larger molecular weight of fluorescent proteins such as EGFP with respect to fluorescein and secondarily to the increased viscosity in the animal tissue compared to the solution: both of them lead to a smaller rate of rotational diffusion that partially preserves the anisotropy induced by photoselection, as we numerically show at the end of this Section. Furthermore, this observation confirms our theoretical model prediction that controlling the illumination polarization (and the consequent photoselection effect) in LSM can maximize the light emitted towards the detector. Finally, the comparable signal level observed for vertical and circular polarizations can originate from this anisotropy since it may lead to a different balance than in the fluorescein solution between the widening of the group of possible target dyes and the decrease in excitation efficiency for the circular polarization.

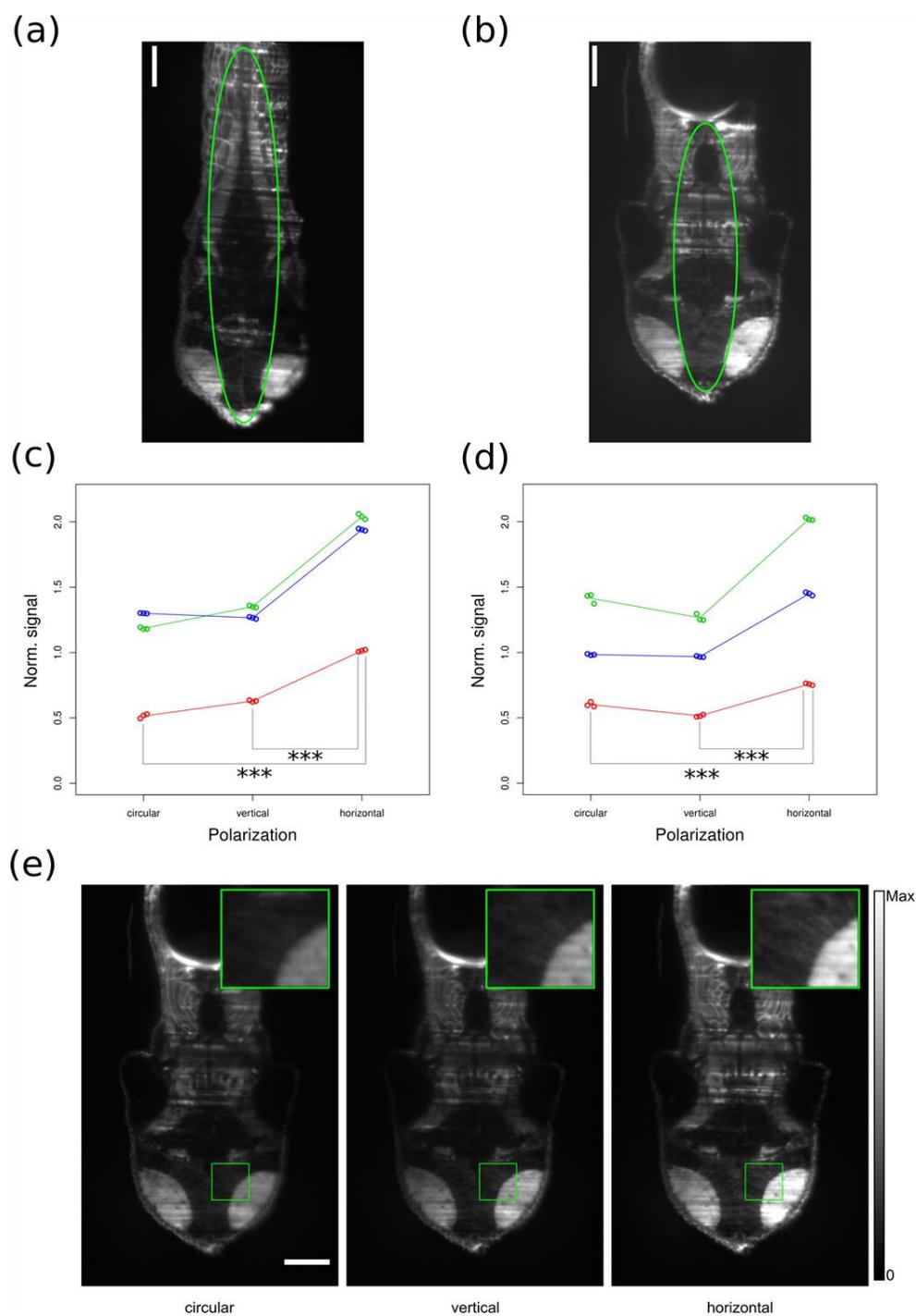

Fig. 3. Imaging of Tg(actin:EGFP) larvae in fixed condition, (a) and (c), and in living condition, (b), (d) and (e). (a), (b) Individual z-slices extracted from volumetric acquisitions of larvae representative of the respective conditions. The green ovals indicate the ROIs traced on these larvae. Scale bars: 100 μm. (c), (d) Scatter plots of the average signal measured from the ROIs as a function of the polarization condition. Each point represents an individual acquisition, the points inherent to the same animal are indicated with the same color in the respective graph. The average values for each animal and for each condition are linked with lines of the same color. In each graph the signal value is normalized to the average of the circular polarization case.

Statistically significant differences are indicated by three asterisks (p-value < 0.0001). (e) Z-slices extracted from volumetric acquisitions of the same larva as in (b) in the three polarization conditions. The signal intensity is mapped as indicated in the grayscale bar on the right. Magnifications of the areas in the green squares are reported in the insets. Scale bar: 100 μm.

Finally, we tested if the examined polarization-dependent effects can be observed also with a fluorescent calcium indicator, such as GCaMP6s. For the fixed-condition, we measured the average fluorescence signal emitted by arbitrarily selected ROIs, similarly to what we did for the EGFP experiments, and we show the results in Figs. 4(a) and 4(c).

In this case as well, we did not observe a significant difference between the circular-polarization condition (14.5 A.U., standard deviation: 1.1 A.U.) and the vertical-polarization (15.5 A.U., standard deviation: 1.3 A.U.) conditions. However, the measured fluorescence levels in the horizontal-polarization condition (26.5 A.U., standard deviation: 1.5 A.U.) showed a large and significant (p-value < 0.0001) increase with respect to the circular-polarization condition (~83%) and the vertical-polarization condition (~71%). These results can be interpreted in the same way as in the EGFP case.

At last, we tested if the GCaMP6s polarization-dependent effects can be observed also during live-imaging. These measures are different with respect to the previous ones, since the cellular calcium levels, and therefore the emitted fluorescence, vary during time, reflecting the time-dependent neuronal activity. In particular, this means that, due to the fluctuations in basal neuronal activity, the fluorescence levels change in the time needed to switch the polarization state. For this reason, we decided to draw ROIs around individual neuronal cells (i.e. the individual sources of the time-dependent signal), and we show the results in Figs. 4(b) and 4(d).

In this case we did not observe a significant difference between the circular-polarization condition (3.6 A.U., standard deviation: 1.9 A.U.) and the vertical- (2.1 A.U., standard deviation: 1.4 A.U.) and the horizontal-polarization (4.8 A.U., standard deviation: 3.5 A.U.) conditions. However, we observed a large (~128.6%) and significant (p-value=0.0016) increase in the fluorescence signal level in the horizontal-polarization condition with respect to the vertical-polarization condition.

The slightly different trends observed for GCaMP6s between the fixed and the living conditions could be ascribed to several factors. The fixation procedure induces cross-linking between molecules that could alter the rotational mobility of the fluorophore. Moreover, the physico-chemical properties of the cytosol change between the living and the fixed states and this medium alteration could affect the motion of the dye. Finally, the fine spatio-temporal biological control of the calcium distribution is completely abolished in the fixed state and therefore the distribution between the bound and the unbound states of the fluorescent sensor is altered in the two cases too, affecting its fluorescence characteristics.

In order to compare the observed results with the theoretical framework described in Section 2, we computed the anisotropy reduction factor $\beta$ and the rotational correlation time $\alpha$ for the different fluorophores, as described in Eqs. (7) and (8) in Section 2. For fluorescein, we set $V \sim 0.34$ nm$^3$ and $\tau \sim 4.1$ ns [43]; for EGFP, we set $V \sim 33$ nm$^3$ [44] and $\tau \sim 2.6$ ns [45] and for GCaMP6s, we set $V \sim 91$ nm$^3$ [44,46] and $\tau \sim 2.5$ ns [47]. Since all measurements were performed at room temperature, we use T=293.15 K and, regarding the medium viscosity, for the water solution we set $\eta \sim 8.9 \cdot 10^{-4}$ Pa· s, while for the intracellular environment we set $\eta \sim 1.2 \cdot 10^{-3}$ Pa· s [44]. With these values, the computed rotational correlation time and anisotropy reduction factor are $\alpha \sim 76$ ps and $\beta \sim 2\%$ for fluorescein, $\alpha \sim 10$ ns and $\beta \sim 79\%$ for EGFP, $\alpha \sim 27$ ns and $\beta \sim 92\%$ for GCaMP6s. While the lifetimes of these three fluorophores are similar, the EGFP and GCaMP6 volumes are one hundred to three hundred times larger than that of fluorescein, predicting a similar increase in $\alpha$ that strongly reduces rotational diffusion and favors the preservation of emission anisotropy induced by photoselection. Our experimental observations are in good agreement with these estimations, since for fluorescein we observed a small residual anisotropy of 3%, which is consistent with $\beta \sim 2\%$; while for EGFP and GCaMP6 we found much larger differences in emission intensity between horizontally and vertically polarized excitation, congruous with the expected anisotropy reduction factors.

We point out the fact that, as we mentioned in Section 1, the populations of the observed fluorophores are here supposed to be randomly oriented. In this situation, linearly-polarized excitation light induces a photoselection on a spatially-isotropic population of dyes and therefore the signal levels are homogeneously affected. On the contrary, in presence of preexistent spatial order of the fluorophores, e.g. due to their binding to spatially-oriented biological structures, polarization-dependent effects on local signal intensity should be expected when using linearly-polarized light. If these effects are not properly taken into account, then they can produce artifacts in the acquired images, such as a relative enhancement or suppression of fluorescence from structures that are co-aligned or orthogonal to the polarization axis. Conversely, the use of circular polarization can bring the benefit of the absence of photoselection, although at the expense of significantly reduced signal intensity in the general case. That is, according to the modeling presented in Section 2, we expect that the local polarization-dependent effects on signal intensity that could arise in presence of spatial order of the fluorophores would be minimized by the use of circular polarization. This could represent a possible advantage of using this type of polarization when observing samples characterized by this particular condition and not aiming for peak signal. Alternatively, a similar result could be achieved also by rotating the polarization plane of the linearly-polarized excitation light to 54.7° (i.e. the "magic angle") with respect to the excited fluorophore population orientation axis [24].

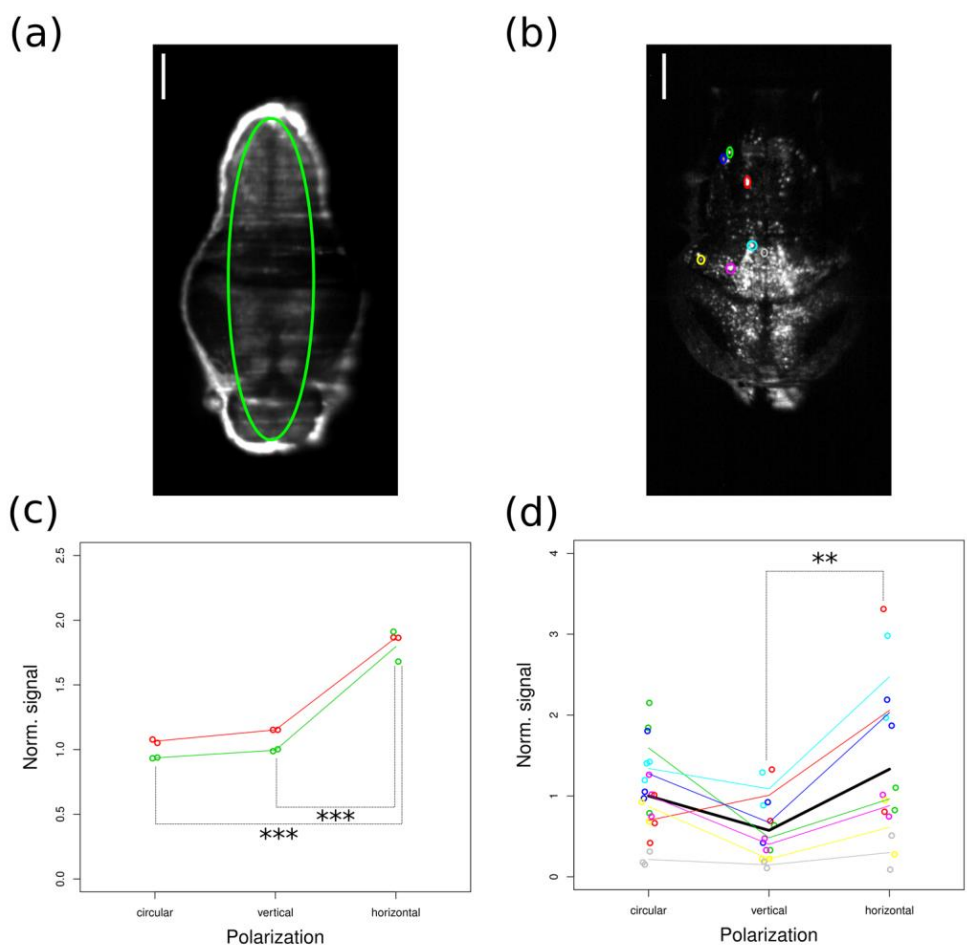

Fig. 4. Imaging of Tg(elavl3:H2B-GCaMP6s) larvae in fixed condition, (a) and (c), and in living condition, (b) and (d). (a) Individual z-slice extracted from the volumetric acquisitions of a representative larva. The green oval indicates the ROI measured for this larva. (b) Maximum projection of a sub-volume of the volumetric stack (70 μm along the dorso-ventral direction from the original 150 μm) of the larva. The colored ovals indicated the different ROIs. Scale bars: 100 μm. (c), (d) Scatter plots of the average signal measured from the ROIs traced on the larvae as a function of the polarization condition. The signal value is normalized to the mean of the circular polarization case. Statistically significant differences are indicated by asterisks (***: p-value < 0.0001, **: p-value = 0.0016). (c) Each point represents an individual acquisition, the points inherent to the same animal are indicated with the same color. The average values for each animal and for each condition are linked with lines of the same color. (d) Different colors indicate different ROIs, as shown in (b). For each color, each point represents an individual acquisition. The average values for each ROI and for each condition are linked with lines of the same color. The thick black line indicates the global averages for each condition.

## 5. Conclusions

In this work we compared the fluorescence signal levels obtained using different excitation light polarization states with fluorescein, EGFP and GCaMP6s fluorophores and different samples in 2P LSM. In all the diverse conditions tested and in agreement with our theoretical expectations, horizontal polarization proved to grant the largest signal levels, while circular polarization generally brought low signal levels. Moreover, vertical polarization gave low fluorescence signal levels with all the biological samples, albeit it provided a large signal for the fluorescein solution. Even if in our observations we focused on functional and structural imaging of animals (zebrafish), we believe that these results apply also to LS cellular imaging [48], since the observed fluorescent signal originates in the cytosol or the nucleosol. Additionally, our experimental findings are directly pertinent for other fluorophores that are randomly-oriented, have similar molecular weights and are embedded in comparable mediums (and thus present close rotational diffusion rates), while for differing conditions other polarization-dependent experimental outcomes should be expected, but still within the landscape provided by our theoretical description. In particular, we expect that the observed signal-enhancement effect would disappear when using very small molecular weight dyes, similarly to what we observed for fluorescein, and increase further when employing fluorophores bound to larger molecular structures. Taken together, these results highlight the importance of controlling the polarization state of the excitation light in 2P LSM of biological samples.

Furthermore, this characterization represents a useful guide to choose the properly-oriented linearly-polarized light when maximization of signal levels is needed. This is particularly important in high-speed 2P LSM, because in this situation (differently from 1P LSM) the acquisition frequency is usually limited by the signal-to-noise ratio and therefore increasing the signal levels is necessary to achieve a higher temporal resolution, by implementing for example strategies that double the frame rate [49].

We anticipate that a polarization-dependence of the signal levels would be present also in 1P LSM (since the involved physical processes are similar in the two modalities), albeit less pronounced due to the reduced photoselection effect, as described in Section 2. Further studies will be necessary to experimentally quantify the exact differences of this effect between 2P and 1P LSM.

**Funding**

European Research Council (692943); H2020 Marie Skłodowska-Curie Actions (793849).

**Acknowledgments**

This project has received funding from the European Research Council (ERC) under the European Union's Horizon 2020 research and innovation programme (grant agreement No 692943).

This project has received funding from the European Union's Horizon 2020 Framework Programme for Research and Innovation under the Marie Skłodowska-Curie grant agreement No. 793849 (MesoBrainMicr).

**Disclosures**

The authors declare no conflicts of interest.